\documentclass[aps,prb,twocolumn,superscriptaddress,amsmath,amssymb,longbibliography]{revtex4-1}
\usepackage[utf8]{inputenc}
\usepackage[english,activeacute]{babel}
\usepackage[T1]{fontenc}
\usepackage[a4paper,vmargin=2.2cm,hmargin=2.0cm]{geometry}
\usepackage{latexsym,amsfonts,amsmath,mathrsfs,amssymb}
\usepackage{braket}
\usepackage[pdftex]{graphicx}
\usepackage{float}
\usepackage{graphicx} 		
\usepackage{hyperref}
\usepackage{multirow}
\usepackage[dvipsnames]{xcolor}
\usepackage{soul}
\usepackage{tabularx}
\usepackage{afterpage}

\newcommand{\rr}{\mathbf{r}}
\newcommand{\sgn}[1]{\text{sgn}\left( #1 \right)}

\newcommand{\sandwich}[3]{\big\langle {#1}  \big\rvert {#2} \big\rvert {#3} \big\rangle}

\begin{document}

\title{Gap engineering and wave function symmetry in C and BN armchair nanoribbons}

\author{Elisa Serrano Richaud}
\affiliation{Universit\'e Paris-Saclay, ONERA, CNRS, Laboratoire d'\'etude des microstructures (LEM), 92322 Ch\^atillon, France}

\author{Sylvain Latil}
\affiliation{Universit\'e Paris-Saclay, CEA, CNRS, SPEC, 91191 Gif-sur-Yvette, France}

\author{Hakim Amara}
\affiliation{Universit\'e Paris-Saclay, ONERA, CNRS, Laboratoire d'\'etude des microstructures (LEM), 92322 Ch\^atillon, France}
\affiliation{Universit\'e Paris Cit\'e, Laboratoire Mat\'eriaux et Ph\'enom\`enes Quantiques (MPQ), CNRS-UMR7162, 75013 Paris, France}

\author{Lorenzo Sponza}
\affiliation{Universit\'e Paris-Saclay, ONERA, CNRS, Laboratoire d'\'etude des microstructures (LEM), 92322 Ch\^atillon, France}
\address{European Theoretical Spectroscopy Facility (ETSF), B-4000 Sart Tilman, Li\`ege, Belgium}

\begin{abstract}
Many are the ways of engineering the band gap of nanoribbons including application of stress, electric field and functionalization of the edges. In this article, we investigate separately the effects of these methods on armchair graphene and boron nitride nanoribbons.
By means of density functional theory calculations, we show that, despite their similar structure, the two materials respond in opposite ways to these stimuli. 
By treating them as perturbations of a heteroatomic ladder model based on the tight-binding formalism, we connect the two behaviours to the different symmetries of the top valence and bottom conduction wave functions.
These results indicate that opposite and complementary strategies are preferable to engineer the gapwidth of armchair graphene and boron nitride nanoribbons.

\end{abstract}

\date{today}

\maketitle
In last decades, graphene (Gr) and hexagonal boron nitride (BN) monolayers have attracted a great deal of interest because of their remarkable transport and optical properties~\cite{Geim_2007, Watanabe_2009, CastroNeto2009, Weng2016, Zhang2017}.
A much explored way to modulate them is by a further reduction of their size like in 2D quantum dots, nanoribbons or nanotubes.
Confinement comes with novel size-dependent features dominated by geometrical parameters and by the characteristics of the edge itself.
This is why nanoribbons are often classified according to their edge shape, which can be zig-zag, armchair, or be structured in a more complex manner~\cite{Ezawa_2006}.
As a matter of fact, the edge characteristics are crucial for the performances of nanoribbon-based devices such as transistors, interconnects and logical devices\cite{Murali2009,Zheng2013,Das2021,MarmolejoTejada_2016,Nishad-2020,Saraswat2021}, photovoltaic applications~\cite{Osella_2012,Saraswat2021}, or chemical sensing\cite{MehdiPour2017,Saraswat2021}.
Gr zig-zag nanoribbons have well localised edge-state which confer them antiferromagnetic properties~\cite{Ezawa_2006,Fujite_1996,Nakada1996,Behzad_2018,Wakabayashi_1999,Yang2007,Son2006}, instead BN zig-zag nanoribbons have an indirect gap and display an intrinsic dipole moment~\cite{Behzad_2018, Nakamura_2005, Park_2008, SHYU_2014,Topsakal_2009,Zhang2008,Wang2011,Jin_2010,Ding_2009}.
At variance, both Gr~\cite{Ezawa_2006,Nakada1996, Behzad_2018,Wakabayashi_1999, Yang2007, Son2006, Barone_2006, JHA_2019, Nishad-2020, Prabhakar_2019, Ren2007,Lu_2009, Raza_2008, SHYU_2014, Sarikavak-Lisesivdin2012,Jippo2013} and BN~\cite{Behzad_2018, Jin_2010, Park_2008,Wang2011, SHYU_2014, Zhang2008,Ding_2009, Topsakal_2009} armchair nanoribbons (AGNR and ABNNR), have no magnetic states and display a direct band gap whose energy depends on the width of the ribbon.
To take full advantage of this richness of properties, several strategies have been explored to engineer the band gap including, among others~\cite{Nakada1996,Osella_2012, Dauber2014,Schwab2015,Nakamura_2005}, application of external electromagnetic fields~\cite{Behzad_2018,Wakabayashi_1999,Park_2008,Zhang2008,Raza_2008,SHYU_2014} of  stress~\cite{Prabhakar_2019,Jin_2010}, or edge functionalization by chemical passivation~\cite{Topsakal_2009,Barone_2006,Ding_2009,Sarikavak-Lisesivdin2012,Jippo2013,JHA_2019,Nishad-2020,Lu_2009,Prabhakar_2019,Zheng2013,Ren2007}.

\begin{figure}[b]
    \centering
    \includegraphics[width=\linewidth]{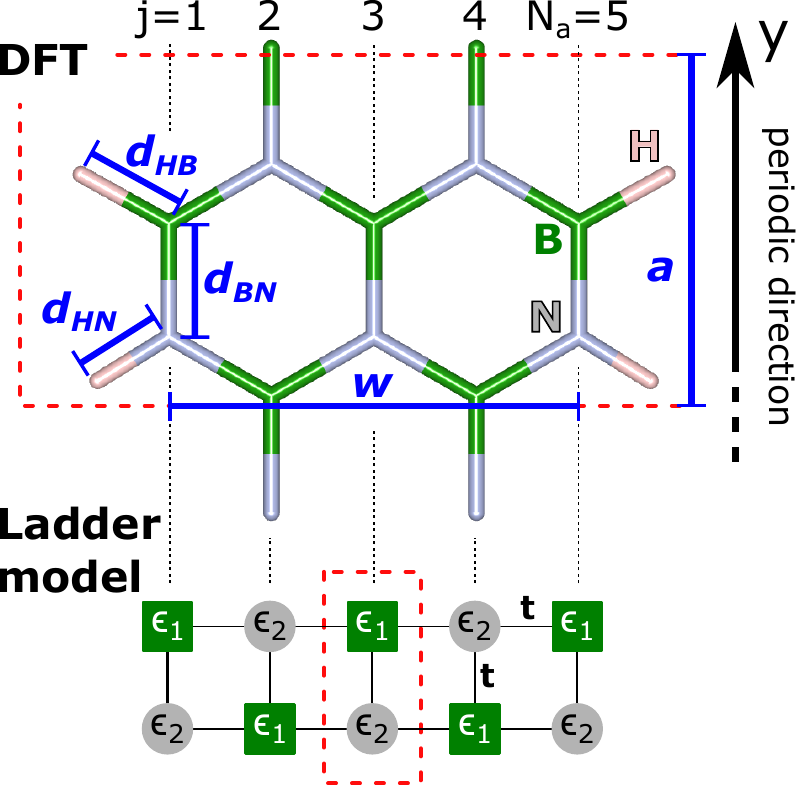}
    \caption{Top: Structure of the reference ABNNR-5.
    Vertical dotted lines enumerate the rows. In blue, we highlight the cell parameter $a$, the width $w$ and the notable lengths $d_{HB}$, $d_{HN}$ and $d_{BN}$. H, N and B atoms are placed respectively in pink, grey and green lattice sites.
    AGNRs are analogously structured with  the relevant edge lengths are labelled $d_{HC}$ and $d_{CC}$.
    Bottom: The corresponding two-atom ladder model. Sites labelled $\mu=1$ are drawn with squares and $\mu=2$ with circles. The color code is the same as in the panel above. The homoatomic ladder model is defined by $\epsilon_1=\epsilon_2=0$.}
    \label{fig:structure}
\end{figure}

In this article, we investigate systematically the response of Gr and BN armchair nanoribbons to these different strategies.
The size of each ribbon is quantified by the width $N_a$ designating the number of longitudinally aligned C-C or B-N dimers as sketched in Figure~\ref{fig:structure}.
To indicate a specific nanoribbon, we append $N_a$ after the type of material, e.g. AGNR-5 indicates an armchair Gr nanoribbon of width $N_a=5$.
Simulations carried out within density functional theory (DFT) include always hydrogens passivating the edges and all structures are fully relaxed, i.e. both in-plane atomic coordinates and the cell parameter $a$ are optimized.
We focus our investigation on the evolution of the gap as a function of $N_a$ upon application of uniaxial stress, biaxial stress and external electric fields.
We also explore the impact of edge functionalization by dividing it into an electrostatic and a deformation component.
For further details on the parameters of the calculations, see Appendix A.

The manuscript is divided as follows:
In part A of Section I, we discuss the variations of the gapwidth upon stimuli corresponding to the different gap engineering strategies simulated at the DFT level.
Successively (part B), we extend to the heteroatomic case a tight-binding ladder model~\cite{Ezawa_2006,Wakabayashi_1999,Son2006} to investigate the response of the two materials.
In Section II, we discuss more in details the specific case of H-passivation solving the ladder model numerically and comparing it with analytical perturbative formulae. 
Conclusions are summarised in Section III.
Appendices A and B are about computational details and some useful mathematical relations.



\begin{figure}
    \centering
    \includegraphics[width=\linewidth]{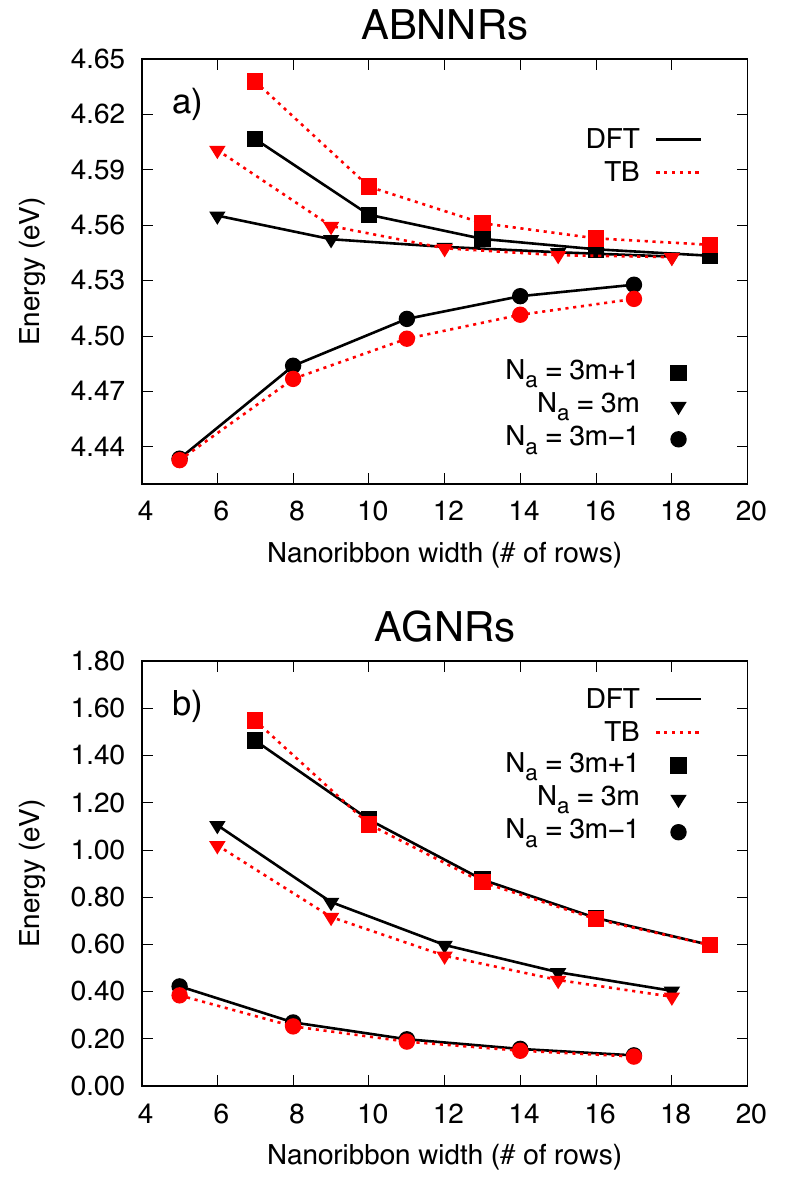}
    \caption{Gap of ABNNRs (a) and AGNRs (b) as a function of $N_a$. Different symbols are associated to different families. Black solid line: DFT calculations. Red dashed line: numerical diagonalization of the ladder model parametrised as in Table~\ref{tab:TB_parameter}. }
    \label{fig:gaps_dft-tb}
\end{figure}

\section{Band gap engineering}

\subsection{Opposite and complementary responses}

\begin{figure*}
    \centering
    \includegraphics[width=0.9\textwidth]{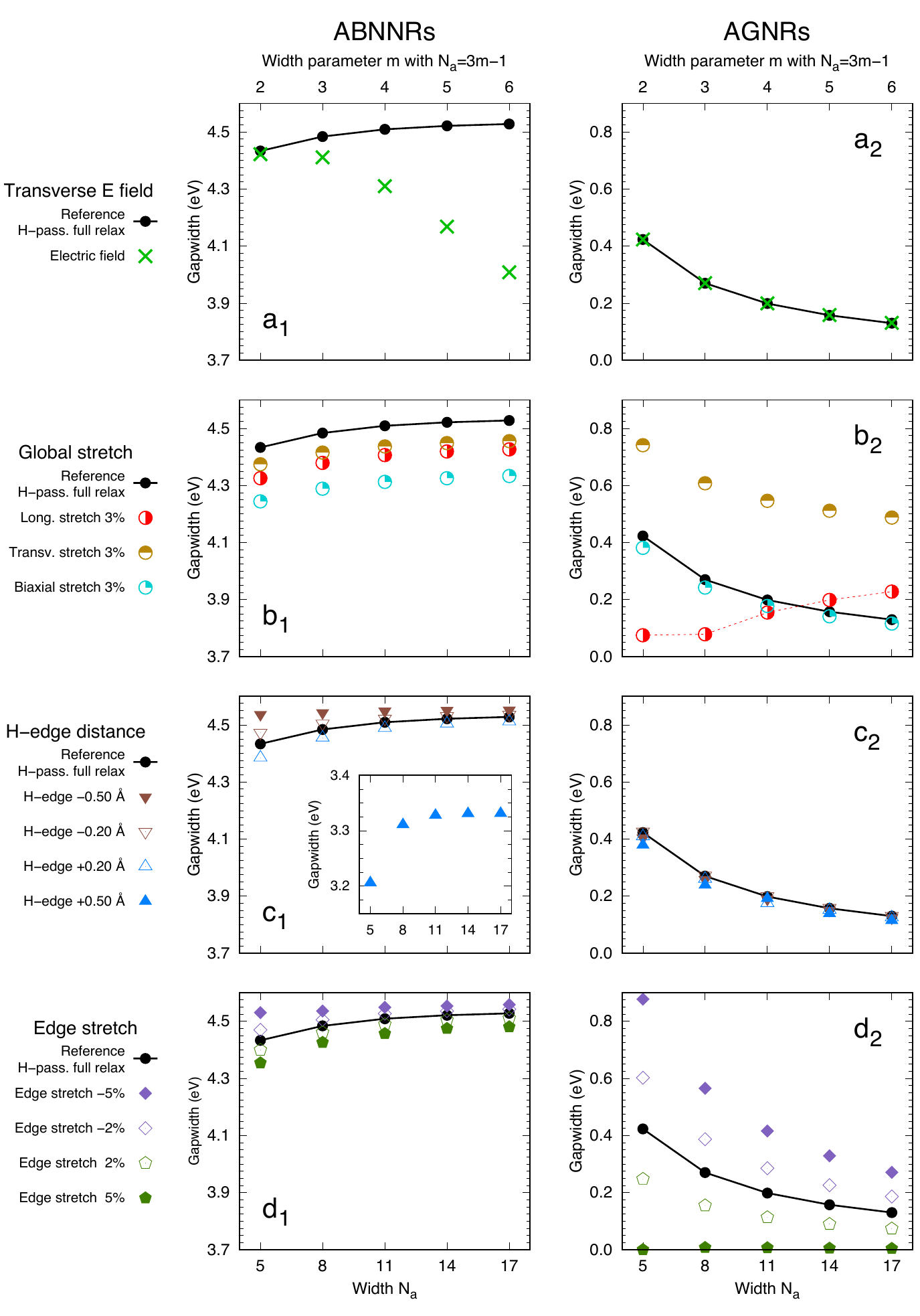}
    \caption{DFT gapwidth of ABNNRs (left panels) and AGNRs (right panels) as a function of $N_a$ under the action of different stimuli accounting for different gap engineering strategies. $a_1$ and $a_2$: transverse electric field; $b_1$ and $b_2$: longitudinal, transversal and biaxial stretch of 3\% (red dashed curve is a gide to the eye); $c_1$ and $c_2$: displacement of the passivating H atoms; $d_1$ and $d_2$: variation of the bondlength at the edges. In all panels, the gapwidth of the reference H-passivated fully relaxed calculation of Figure~\ref{fig:gaps_dft-tb} is reported with a bulleted solid black line.} 
    \label{fig:check_model}
\end{figure*}

\begin{figure*}
    \centering
    \includegraphics[width=0.9\textwidth]{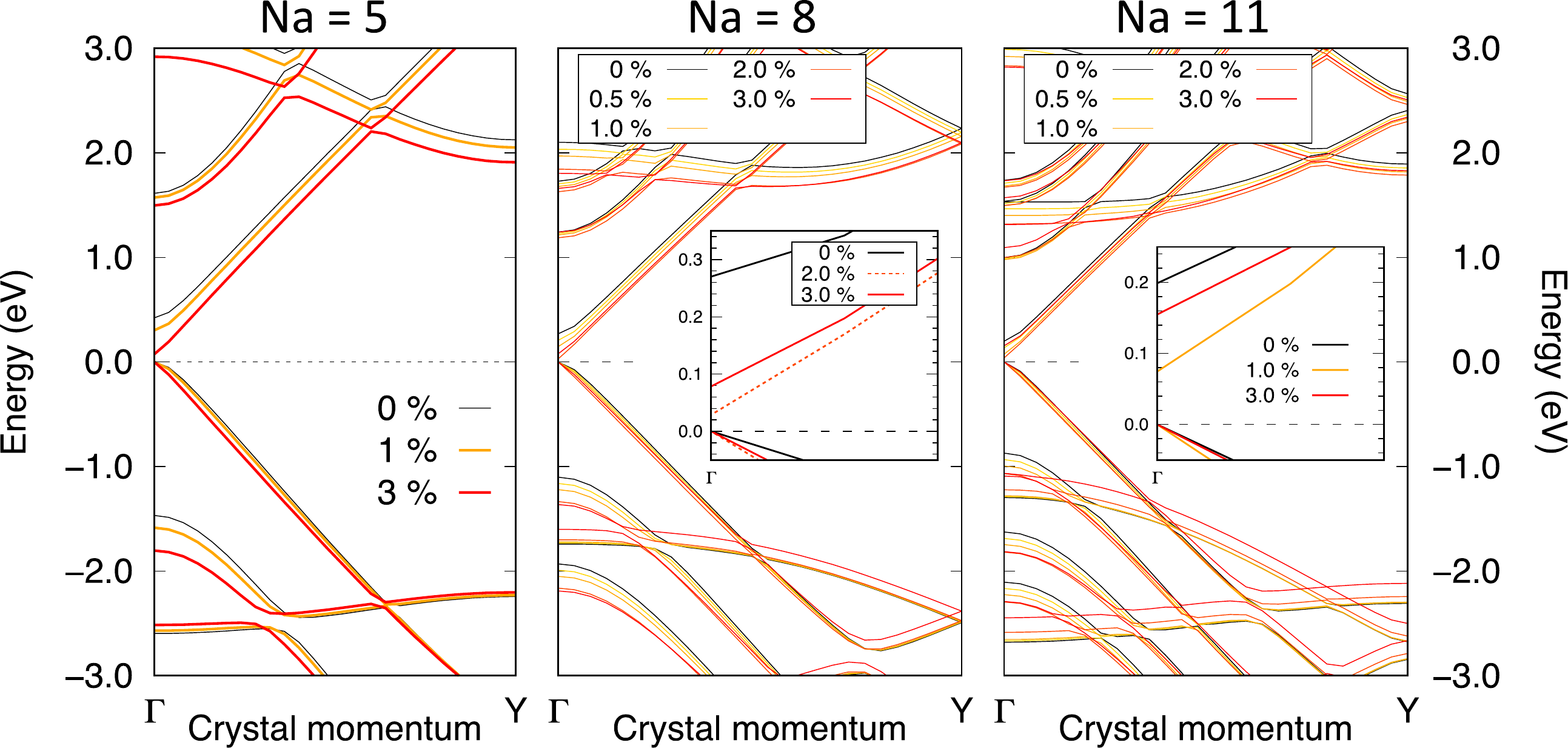}
    \caption{DFT band structure of AGNRs with $N_a=5, 8$ and $11$ sustaining various longitudinal stresses.}
    \label{fig:band_structure_bouncing}
\end{figure*}

The electronic structure of BN and Gr nanoribbons has been thoroughly studied in the past~\cite{Ezawa_2006,Fujite_1996,Nakada1996,Behzad_2018,Wakabayashi_1999,Yang2007,Son2006,Behzad_2018, Nakamura_2005, Park_2008, SHYU_2014,Topsakal_2009,Zhang2008,Wang2011,Jin_2010,Ding_2009,Ezawa_2006,Nakada1996, Behzad_2018,Wakabayashi_1999, Yang2007, Son2006, Barone_2006, JHA_2019, Nishad-2020, Prabhakar_2019, Ren2007,Lu_2009, Raza_2008, SHYU_2014, Sarikavak-Lisesivdin2012,Jippo2013,Behzad_2018, Jin_2010, Park_2008,Wang2011, SHYU_2014, Zhang2008,Ding_2009, Topsakal_2009}.
Our DFT band gap of ABNNRs and AGNRs for width ranging from $N_a=5$ to $N_a=19$ are reported in red in the two panels of Figure~\ref{fig:gaps_dft-tb} and are in good agreement with similar calculations in literature.
We recall that, owing to the 1D confinement, 
the gapwidth of both materials falls in one of the three families $N_a = 3m-1$, $N_a=3m$ or $N_a=3m+1$ (with $m\in \mathbb{N}*$).
The ordering of the families is the same in the two materials, with the $N_a = 3m -1$ branch always lowest in energy.
We remark that size effects on thinner ribbons can be quite large: in ABNNRs the gap variation is roughly $\pm 0.1$~eV with respect to the asymptoptic limit, which happens to be lower than in the isolated monolayer because of a residual edge contribution~\cite{Park_2008,Topsakal_2009}.
In AGNRs size effects are even stronger, the gap opens up to 1.6 ~eV at width $N_a = 7$ and ranges between 0.1 and 0.4 in the $N_a=3m-1$ family.

\bigskip

We can now proceed with the inclusion of stimuli accounting for different gap engineering strategies.
For reasons that will appear clearer later, we focus our analysis on the family $N_a=3m-1$.
In Figure~\ref{fig:check_model}, we report the gapwidth of ABNNRs (left panels) and AGNRs (right panels) belonging to this family as a function of $N_a$ and under the effect of different stimuli. 
In all panels, the bulleted black solid line are the reference band gaps of the fully-relaxed calculations (those reported in Figure~\ref{fig:gaps_dft-tb}).
Before discussing the details, one result immediately jumps out.
The two materials display opposite and complementary responses. 
When ABNNRs show a violent modification of their gapwidth, AGNRs show negligible variations and the other way around.

In literature, the effect of a transversal (along $x$) electric field is predict to close the gap of ABNNRs already at weak intensity, with stronger variations in wider ribbons~\cite{Park_2008,Behzad_2018,Zhang2008}.
Instead, in AGNRS a width-dependent threshold intensity must be passed below which the gap is constant~\cite{Behzad_2018,Raza_2008}.
Our results are reported in panels $a_1$ and $a_2$ with green crosses for a field of 0.5 V/nm, corresponding to a very weak field.
For the comparison to be meaningful, the field is the same in the two materials.
In agreement with literature, DFT predicts a relatively violent gap closing in ABNNRs where it decreases from 4.4~eV to 4.0~eV eV with an approximately linear trend, whereas the gap of AGNRs is basically not affected. 

The effect of global stretch is reported in panels $b_1$ and $b_2$. In addition to the reference values, we report the band gap of ribbons experiencing the same longitudinal ($y$), transversal ($x$) or biaxial stretch.
More precisely, a tensile strain of 3\% is applied to parameters $a$, $w$ or both.
This is a very low strain, corresponding to a harmonic elastic regime of deformation~\cite{Topsakal_2009}, but what is most important is that it is the same value for all calculations. 
At variance from before, here DFT predicts that the gap of ABNNRs undergoes weaker variations than that of AGNRs.
Typically, uniaxial variations in ABNNRs are of about 0.1~eV, whereas they reach about 0.3~eV in AGNRs.
A comparison with literature is not simple, but for instance a stress of 7\% produces gap variations of about 0.3 eV in ABNNR-7~\cite{Jin_2010} and more than the double in AGNR-7~\cite{Prabhakar_2019}.
Interestingly, a biaxial stress has almost no effect on AGNRs, as if its $y$ and $x$ elongation components cancel each other, contrary to what is observed in ABNNRs where the two seem to add in closing the gap by about 0.2 eV.
Besides this, a peculiar trend is observed in longitudinally stretched AGNRs, highlighted by a red dashed line.
This is related to a sort of ``\emph{bouncing}'' of the band gap that closes in small ribbons and opens again in ribbons larger than the characteristic width $N_a \approx 11$.
This effect is also visible in the data by Prabhakar and coworkers\cite{Prabhakar_2019} in the case of  AGNR-7.
This ``\emph{bouncing}'' behaviour can be appreciated more clearly in Figure~\ref{fig:band_structure_bouncing} where we report the AGNRs band structure in the vicinity of the gap for different widths and values of the longitudinal stretch. 
The higher the stretch, the smaller the critical width. 

Edge functionalization is actually a very complex mechanism which may involve global and edge strain, variations of the electrostatic potential felt by all atoms of the ribbon and some charge transfer.
Having already addressed global strain, here we concentrate the study on the electrostatic variation (panels $c$) and edge strain (panels $d$).
These separate effects have been studied in the past in the case of AGNRs~\cite{Son2006,Lu_2009}, here we extend the analysis to ABNNRs and make a quantitative comparison by considering the two contributions separately.
In panels $c_1$ and $c_2$, we report the gap under variations of $d_{HX}$ of $\pm0.2$ or $\pm0.5$ \AA{ } ($X=B$, $N$ or $C$).
Here $+$ signs indicate a displacement away from the edge, $-$ signs toward it.
Notice that the asymptotic limit of the ABNNRs gapwidth when H atoms are moved 0.50 away from the edges, is not the reference gap, but a sizeably smaller one much closer to that of the non-passivated ribbon~\cite{Topsakal_2009}.
Finally, in panels $d_1$ and $d_2$ we report the band gap under expansion ($+$) and contraction ($-$) of $d_{XY}$ (with $XY=BN$ or $CC$) of 2\% or 5\%.
Considered together, these four panels show that edge functionalization modifies efficiently the gap of both materials, though through different mechanisms.
While ABNNRs respond mostly to variations of the electrostatic potential, AGNRs are mostly sensitive to the induced strain at the edges, in agreement with literautre\cite{Son2006,Lu_2009}. 
In both cases, gap variations decrease for increasing width, as expected in the limit of isolated monolayer (infinite width limit).

\begin{figure}[b]
    \centering
    \includegraphics[width=\linewidth]{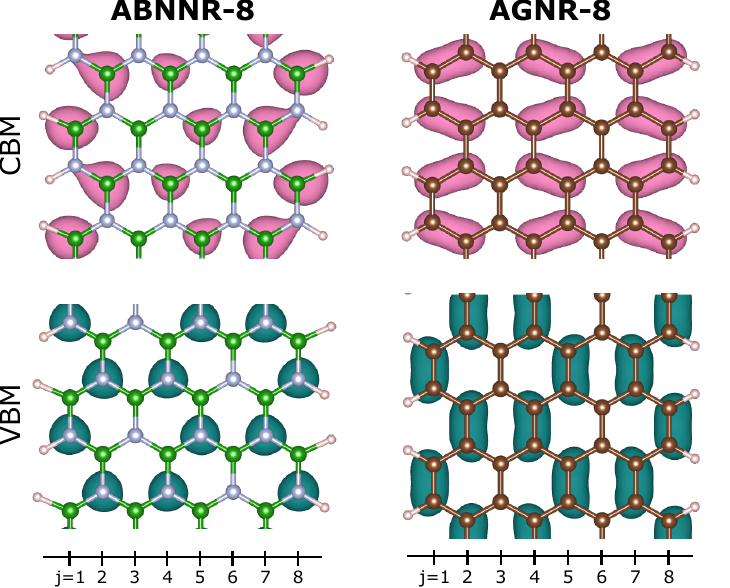}
    \caption{Reference DFT charge density of the CBM  (pink - top panels) and the VBM (dark green - bottom panels) in ABNNR-8 (left panels) and AGNR-8 (right panels).
    B atoms are placed n green sites, N atoms in grey sites and C atoms in brown sites. Passivating H atoms are the pink spheres on the edges.}
    \label{fig:wavefunctions}
\end{figure}

Let us now stop here and summarise our results.
We have shown that the gapwidth of ABNNRs is strongly sensitive to variations of the electrostatic environment, indicating the application of electric fields as the most effective way to engineer the gap of ABNNRs.
Edge functionalization can be effective on the condition that the electrostatic potential is strongly modified by the passivating species.
On the other hand, AGNRs are demonstrated to respond more strongly to mechanically applied uniaxial strain or to the edge strain induced by the functionalization.
Such an opposite and complementary response must be rooted in the impact that different stimuli have on the valence band maximum (VBM) and the conduction band minimum (CBM).
As a matter of facts, these states have very different localization characteristics in the two materials.
While in ABNNRs, the states localize basically on the lattice sites like in the monolayer\cite{Galvani2016}, in AGNRs they are spread along the C-C bonds. 
As a clear example of this, in Figure~\ref{fig:wavefunctions} we report the partial density $|\psi(\rr)|^2$ in $\Gamma$ of the last occupied and lowest empty states in $N_a=8$ reference ribbons.

In the next part, we develop a two-atom ladder model to investigate further the consequences of this difference.



\subsection{Two-atom ladder model}
Here we employ the tight-binding \emph{ladder} model which has been initially introduced to study the gap of AGNRs\cite{Wakabayashi_1999,Fujite_1996,Nakada1996,Ezawa_2006,Son2006}.
We extend it to the heteroatomic case with the intent of describing both Gr and BN nanoribbons' gap.
The resulting heteroatomic ladder model is sketched in the bottom panel of Figure~\ref{fig:structure}.
Electric fields, global strain and edge functionalization decomposed in its electrostatic and edge-strain components are introduced through first order perturbation theory via changes of the on-site and hopping parameters of the model, as suggested in~\cite{Ezawa_2006,Son2006}.




Let us first diagonalize the unperturbed Hamiltonian which reads:
\begin{equation}
H^0 = \sum_{j,\mu} \left( \epsilon_{\mu}  \ket{\mu, j}  + \sum_{\text{n.n.}} t\ket{\mu', j'} \right) \bra{\mu ,j}\,.
\end{equation}
The index $j\in\left[1,N_a\right]$ labels the row, while $\mu=1,2$ indicates both the atomic site and the atomic species in the row ($C_1$ or $C_2$ in AGNRs and $B$ or $N$ in ABNNRs). 
Note that $\mu=1$ is below $\mu=2$ in even rows, and vice-versa in odd rows.
The basis function $\braket{\rr |\mu, j}$ is the $p_z$ orbital of the atom $\mu$ of the $j$th row.
The inner sum is limited to first neighbors and the onsite energies are $\epsilon_{1}=-\epsilon_2=\epsilon$.
 In ABNNRs, $\epsilon>0$ to associate N atoms a negative on-site energy and B atoms a positive one\cite{Galvani2016}.
 AGNRs are obtained for $\epsilon=0$.

The discrete spectrum of $H^0$ is $E_{n\pm}^0 = \pm \mathcal{E}_n$ where $n\in[1, ... , N_a]$ and the following definitions are introduced: $\mathcal{E}_n=\sqrt{ \epsilon^2 + \tau^2_n  }$ and $\tau_n = t \left[ 1 + 2 \cos\left( \theta_n \right) \right]$ and  $\theta_n = n\pi/(N_a +1)$.
The $-$ and $+$ sings label occupied and empty states respectively, which implies that the valence band maximum (VBM) and the conduction band minimum (CBM) are those that minimize $|\tau_n|$.
They are labelled $\tilde{n}=2m+1$ in families $N_a=3m$ and $N_a=3m+1$, and  $\tilde{n}=2m$ if $N_a=3m-1$.
The generic normalised unperturbed eigenstate $\ket{n,\pm}$ reads
\begin{equation}\begin{split}
\ket{n,\pm} &=\sqrt{\frac{2}{N_a+1}} \sum_{j=1}^{N_a} \sin\left( j \theta_{n} \right) \times \\ 
&\qquad \times \Big[ D^{n,\pm}_1 \ket{1,j} +  D^{n,\pm}_2 \ket{2,j}\Big]
\label{eq:wavefunction_generic}
\end{split}\end{equation}
with the coefficients
\begin{equation}
\begin{split}
D^{n\pm}_1 &= - \frac{\tau_n}{\sqrt{2\mathcal{E}_n ( \mathcal{E}_{n} \mp \epsilon)}}  \quad\text{and}\\  D^{n\pm}_2 &= \frac{\mp \mathcal{E}_n + \epsilon}{\sqrt{2\mathcal{E}_n ( \mathcal{E}_{n} \mp \epsilon)}}\,.
\end{split}
\label{eq:coefficients_generic}
\end{equation}
Finally, the unperturbed band gap of the ladder model reads
\begin{equation}
    E_g^0 = \left\{ \begin{array}{ll}
    2\epsilon & \text{for } N_a = 3m-1\\
    2\mathcal{E}_{2m+1} & \text{for the other values of $N_a$} \end{array}  \,.  \right.
    \label{eq:gap_unperturbed_general}
\end{equation}

\begin{figure}
    \centering
    \includegraphics[width=1.0\linewidth]{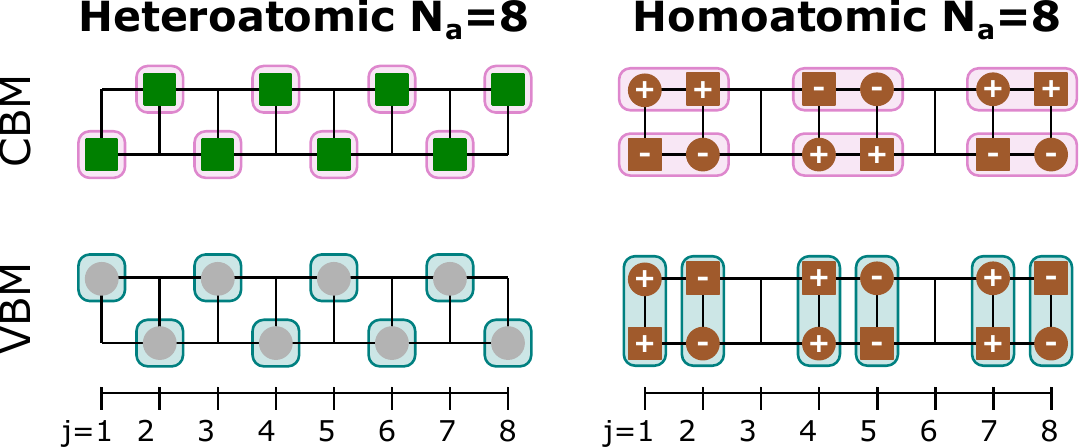}
    \caption{Representation of the last occupied (VBM) and first unoccupied (CBM) eigenstates of the ladder model.
    Squares and circles indicate sites with label $\mu=1$ or $\mu=2$ respectively.
    Lacking symbols designate sites with a vanishing associated total weight in expression~\eqref{eq:wavefunction_generic}.
    In the heteroatomic model, all weights are either $\sqrt{3}/2$ or 0.
    In the homoatomic one, the $\pm$ symbols are for weights of $\pm \sqrt{3/8}$.
    Bonding combinations are formed between sites sharing the same sign.
    Shaded areas highlight the similarity with Figure~\ref{fig:wavefunctions}.}
    \label{fig:wfc_tb}
\end{figure}


It is worth stopping here to evaluate explicitly the wavefunctions of the VBM and CBM.
\begin{description}
\item[ABNNRs] Since $\tau_{2m}\equiv0$, in the $N_a=3m-1$ family the Hamiltonian of the VBM and CBM is effectively a non interacting Hamiltonian. 
The corresponding wave function coefficients read:
\begin{eqnarray}
\text{VBM:} & D_1^{2m,-} = 0 &\quad\text{and}\quad D_2^{2m,-} = 1 \label{eq:coefficients_BN_3m-1_valence}\\
\text{CBM:} & D_1^{2m,+} = 1 &\quad\text{and}\quad D_2^{2m,+} = 0\label{eq:coefficients_BN_3m-1_conduct}
\end{eqnarray}
corresponding to pure $N$ states in valence ($\mu=2)$, and pure $B$ states in conduction ($\mu=1$), as expected in all BN-based materials\cite{Galvani2016}.
These coefficients have to be multiplied by the standing wave envelop $\sin\left( j \theta_{2m}\right)$.
The example for $N_a=8$ is given in left panels Figure~\ref{fig:wfc_tb} to be compared with the partial charge densities reported in Figure~\ref{fig:wavefunctions}.
The ladder model results match extremely well with the first-principle calculations.
In the other two families, $|\tau_{2m+1}|$ does not vanish exactly, but is minimized, so the deviation from perfectly pure states is small and actually gets smaller for wider ribbons. 

\item[AGNRs] In the homoatomic case, one has to set $\epsilon=0$ in the Hamiltonian prior to any further manipulation.
This condition changes completely the nature of the wave functions, leading to symmetric and antisymmetric combinations of the atomic-centered wave functions. 
Regardless of the gap family, the coefficients read
\begin{eqnarray}
\text{VBM:} & D_1^{\tilde{n},-} = -\frac{\sgn{\tau_{\tilde{n}}}}{\sqrt{2}} &\quad\text{and}\quad D_2^{\tilde{n},-} = \frac{1}{\sqrt{2}} \label{eq:coefficients_Gr_valence}\\
\text{CBM:} & D_1^{\tilde{n},+} = -\frac{\sgn{\tau_{\tilde{n}}}}{\sqrt{2}} &\quad\text{and}\quad D_2^{\tilde{n},+} =-\frac{1}{\sqrt{2}}\,.\label{eq:coefficients_Gr_conduct}
\end{eqnarray}
Here $\sgn{x}$ takes value $1$ if $x>0$ and $-1$ is $x \leq 0$.
The final wavefunction at $N_a=8$, including the envelop prefactors, is reported in the right panels of Figure~\ref{fig:wfc_tb} and once again the match with DFT partial charges of Figure~\ref{fig:wavefunctions} is excellent.
Note that in the $N_a=3m-1$ family, it is incorrect to distinguish between VBM and CBM states within the model, because they are actually degenerate~\cite{Son2006} at $E_{2m,\pm}^0=0$.
One should rather speak about symmetric (CBM) and antisymmetric (VBM) states, but we will keep the VBM and CBM labelling for convenience.

\end{description}


Now that we solved the unperturbed problem, we can follow the suggestions of~\cite{Ezawa_2006,Son2006} and introduce the different band engineering manipulations through a perturbation Hamiltonian $\delta H$ which includes:\\
\begin{itemize}
\item Transverse electric field terms $j \cdot f_\mu$ with $\mu=1,2$.
The $f_\mu$ perturbative parameters modify the on-site energies of the entire ribbon linearly with the $x$ coordinate of the atomic site, here replaced by the row index $j=1,...,N_a$;
\item Longitudinal ($y$) and Transverse ($x$) stretch terms $\delta t_\parallel$ and $\delta t_\perp$ respectively, added to all hopping terms depending on their direction.
\item Edge electrostatic corrections $\delta \epsilon_{\mu}$ modifying the on-site energies of atoms at the edges (i.e. only on $j=1$ and $j=N_a$); 
\item Edge stress terms $\delta t_e$ modifying the hopping between atoms forming the edges, i.e. belonging to rows $j=1$ and $j=N_a$.
\end{itemize}
This perturbative Hamiltonian diagonalizes on the basis as
\begin{equation}
\sandwich{\mu,j}{\delta H}{\mu,i}  =  \left[ j f_\mu  + \delta\epsilon_\mu \left( \delta_{j,1} + \delta_{j,N_a} \right) \right] \, \delta_{j,i}
\end{equation}
for both $\mu=1$ and $\mu=2$; 
and
\begin{equation}
\begin{split}
\lefteqn{ \sandwich{2,j}{\delta H}{1,i}  =  \sandwich{1,j}{\delta H}{2,i}  = } \\
&\qquad = \left[\delta t_\parallel + \delta \tau \left(\delta_{j,1} + \delta_{j,N_a} \right)\right] \delta_{j,i} +\\
&\qquad\quad+ \delta t_\perp \left( \delta_{j-1,i} +  \delta_{j+1,i} \right) \,.    
\end{split}
\end{equation}
The resulting correction to the generic state $\sandwich{n,\pm}{\delta H}{n,\pm} = 
\delta F_{n\pm} + 
\delta S_{n\pm} + \delta C_{n\pm}$ is the sum of three perturbations corresponding to the application of a transverse electric field ($\delta F$), the application of uniaxial and/or biaxial stress ($\delta S$) and to edge functionalization  ($\delta C$).
After some not so obvious trigonometric manipulation reported in Appendix (Section~\ref{sec:appendix_trigonometric}), one gets:
\begin{equation}
 \delta F_{n\pm} =\frac{N_a+1}{2} \Big[ \left(D_1^{n\pm}\right)^2  f_1  +  \left(D_2^{n\pm}\right)^2  f_2 \Big] \,, \label{eq:field_term}   
\end{equation}
\begin{equation}
\delta S_{n\pm} = 2D_1^{n\pm} D_2^{n\pm} \left[ \delta t_\parallel + 2\cos(\theta_n)\delta t_\perp\right]
\label{eq:stress_term}
\end{equation}
and 
\begin{equation}
\begin{split}
\delta C_{n\pm} &= \frac{4 \sin^2(\theta_n)}{N_a + 1}  \Big[ 2D_1^{n\pm}D_2^{n\pm} \delta t_e \,+ \\ &\quad+ \left(D_1^{n\pm}\right)^2 \delta \epsilon_1 + \left(D_2^{n\pm}\right)^2 \delta \epsilon_2 \Big]\label{eq:chemistry_term}    
\end{split}
\end{equation}

\begin{widetext}
The perturbative correction to the band gap is $\delta E_g= \sandwich{\tilde{n},+}{\delta H}{\tilde{n},+}-\sandwich{\tilde{n}, - }{\delta H}{\tilde{n}, - }$. Note however that in the homoatomic $N_a=3m-1$ family, since the unperturbed solution is gapless, the absolute value of this expression must be taken.
By plugging into this expression the terms~(\ref{eq:chemistry_term}~-~\ref{eq:stress_term}), one gets the generic gap correction
 \begin{equation}\begin{split}
 \delta E_g &= \Big[ \left(D_1^{\tilde{n}+}\right)^2 - \left(D_1^{\tilde{n}-}\right)^2 \Big] \Bigg[  \frac{4 \sin^2(\theta_{\tilde{n}})}{N_a + 1}   \delta \epsilon_1 +  \frac{N_a+1}{2} f_1   \Bigg] + \Big[ \left(D_2^{\tilde{n}+}\right)^2 -   \left(D_2^{\tilde{n}-}\right)^2\Big]  \Bigg[  \frac{4 \sin^2(\theta_{\tilde{n}})}{N_a + 1}   \delta \epsilon_2 + \frac{N_a+1}{2} f_2  \Bigg] +\\
 & \quad +  2\Big[ D_1^{\tilde{n}+} D_2^{\tilde{np}+} - D_1^{\tilde{n}-} D_2^{\tilde{n}-} \Big]  \left[ \frac{4\sin^2\left(\theta_{\tilde{n}}\right)}{N_a+1}\delta t_e + \delta t_\parallel +  2\cos(\theta_{\tilde{n}}) \delta t_\perp \right]\,.
 \end{split}
 \label{eq:perturbed_gap_all_general}
 \end{equation}
We finally understand why the two materials have different responses to the different stimuli. 
In fact, if we now insert the wave function coefficients~\eqref{eq:coefficients_generic} into~\eqref{eq:perturbed_gap_all_general}, one finds different band gap corrections for ABNNRs and AGNRs.
\begin{description}
\item[ABNNRs] In the heteroatomic solution, for families $N_a=3m$ and $N_a=3m+1$, one get
\begin{equation}
\begin{split}
\delta E_g &= \frac{\epsilon}{\mathcal{E}_{2m+1}} \Bigg[  \frac{4 \sin^2(\theta_{2m+1})}{N_a + 1}  \left(  \delta \epsilon_1  - \delta \epsilon_2\right) + \frac{N_a+1}{2} \left( f_1 - f_2 \right)  \Bigg] \, +\\
& \quad + \frac{2\tau_{2m+1}}{\mathcal{E}_{2m+1}}  \Bigg[  \frac{4 \sin^2(\theta_{2m+1})}{N_a + 1} \delta t_e + \delta t_\parallel +  2\cos(\theta_{2m+1}) \delta t_\perp \Bigg] \,.
\end{split}
\label{eq:perturbed_gap_heteroatomic_generic}
\end{equation}
Notice that, since $|\tau_{\tilde{n}}|$ is minimized in the VBM and CBM and it gets closer to 0 as the width increases, the gap correction is dominated by the first term containing only on-site perturbations, i.e. the  application of electric fields and the electrostatic component of the edge functionalization.
The extreme case of this is encountered in the $N_a=3m-1$ family, where $\tau_{\tilde{n}}\equiv 0$ and the wave function coefficients are the pure state combinations reported in expressions~\eqref{eq:coefficients_BN_3m-1_valence} and~\eqref{eq:coefficients_BN_3m-1_conduct}.
This makes disappear completely the second term and the resulting variation of the gap reads simply
\begin{equation}
\delta E_g =   \frac{1}{m} \left( \delta \epsilon_1  - \delta \epsilon_2 \right) + \frac{3}{2}m \left( f_1 - f_2 \right)  \, .
\label{eq:perturbed_gap_heteroatomic_3m-1}
\end{equation}
\item[AGNRs]
In all famlilies of the homoatomic case, the wave function coefficients are the symmetric and antisymmetric combinations~\eqref{eq:coefficients_Gr_valence} and~\eqref{eq:coefficients_Gr_conduct}.
As a consequence, all the terms modifying on-site energies vanish because of the mutual cancellation of the coefficients $\left(D^{\tilde{n}+}_\mu\right)^2 - \left(D^{\tilde{n}-}_\mu\right)^2$ in~\eqref{eq:perturbed_gap_all_general}.
Instead the coefficients of the global and edge stress terms add together constructively into $D^{\tilde{n}+}_1D^{\tilde{n}+}_2 - D^{\tilde{n}-}_1D^{\tilde{n}-}_2 = 1$.
In families $N_a=3m-1$ and $N_a=3m$, this leads to the perturbative gap correction
\begin{equation}
\delta E_g =  \frac{8 \sin^2(\theta_{2m+1})}{N_a + 1} \delta t_e  + 2\delta t_\parallel +  4\cos(\theta_{2m+1}) \delta t_\perp  \, , 
\label{eq:perturbed_gap_homoatomic_generic}
\end{equation}
whereas in the $N_a=3m-1$ family, the final expression reads
\begin{equation}
\delta E_g =  2  \left| \frac{1}{m} \delta t_e + \delta t_\parallel - \delta t_\perp \right|\,.
\label{eq:perturbed_gap_homoatomic_3m-1}
\end{equation}

\end{description}
\end{widetext}

Expressions~\eqref{eq:perturbed_gap_heteroatomic_3m-1} and~\eqref{eq:perturbed_gap_homoatomic_3m-1} indicate how the band gap varies in the family $N_a=3m-1$ and can be qualitatively compared with the DFT results of Figure~\ref{fig:check_model}.
Concerning the applied field engineering method (panels $a_1$ and $a_2$) the model predics correctly the strong dependence of ABNNRs on the electric field together with its linear dependence on $N_a$.
In the case of applied stress (panels $b_1$ and $b_2$), the perturbative formulae predict the right qualitative behavior with the AGNRs much more sensitive to the ABNNRs (for which there is no variation in the model to be compared with the relatively weak variations predicted by DFT).
Also well captured by the model is the opposite sign of $\delta t_\parallel$ and $\delta t_\perp$ in equation~\eqref{eq:perturbed_gap_homoatomic_3m-1} which indicates two opposite results upon application of  longitudinal and perpendicular stress.
Even though the gap ``bouncing" discussed earlier can not be described by our TB model, the opposite sign predicts the negligible effect of biaxial strain.
Coming to the two components of the edge functionalization, the responses predicted by the model are again in very good agreement with the DFT results.
According to the formalae, ABNNRs are extremely sensitive to on-site modifications of this gap engineering method, that is to the passivation-induced electrostatic variations (panels $c_1$ and $c_2$) contrary to AGNRs that are instead more sensitive to the hopping variations, i. e. to the edge strain (panels $d_1$ and $d_2$). 

The origin of the opposite and complemetary response in ABNNRs and AGNRs is now made transparent by the heteroatomic ladder model and is ultimately related to the symmetries of the VBM and CBM in the two materials. 
In ABNNRs they are basically pure B or N states, confering the ribbon a higher sensitivity to on-site variations.
In AGNRs, on the contrary, they are basically symmetric and antisymmetric combinations which make the wave functions extend on the bonding. This makes AGNRs much more sensitive to changes of the hopping parameters.

\section{Ladder model of H-passivated nanoribbons}

\begin{table}[b]
    \centering
    \begin{tabular}{c | c | c | c }
        \multicolumn{4}{c}{ABNNR} \\ \hline\hline
        $\epsilon_B = -\epsilon_N$ & $t$ & $\delta \epsilon$ & $\delta t_e$ \\ 
         2.285 & -2.460 & -0.145 & -0.190
    \end{tabular}
    
    \vspace{0.3cm}
    
    \begin{tabular}{c | c | c | c }
        \multicolumn{4}{c}{AGNR} \\ \hline\hline
        $\epsilon_{C_1} = \epsilon_{C_2}$ & $t$ & $\delta \epsilon$ & $\delta t_e$\\
        0.0 & -2.600 & 0.0 & -0.400
    \end{tabular}
    \caption{TB parameters in eV for H-passivated nanoribbons. Resulting gaps are reported in Figure~\ref{fig:gaps_dft-tb}. }
    \label{tab:TB_parameter}
\end{table}

In this section, we provide some additional information on the ladder model when applied to H-passivated ribbons. 
We retain only the edge-functionalization parameters of $\delta H$, nalemy $\delta \epsilon_\mu$ and $\delta t_e$ and we diagonalize numerically the Hamiltonian $H=H^0+\delta H$.
Successively we fit the parameters of $H$ against the DFT band gaps reported in Figure~\ref{fig:gaps_dft-tb}.
Guided by physical intuitions we take $\delta \epsilon_1=\delta \epsilon_2=\delta \epsilon$ in the case of AGNRs, and $\delta \epsilon_1=-\delta \epsilon_2 = \delta \epsilon$ in the case of ABNNRs.
The resulting parameters and the TB band gap are reported respectively in Table~\ref{tab:TB_parameter} and in Figure~\ref{fig:gaps_dft-tb} with red dashed lines.
Note that we take a non-vanishing value for $\delta t_e$ in ABNNRs even though its influence should be weak.
This is indeed the case in the $N_a=3m-1$ family (expression~\eqref{eq:perturbed_gap_heteroatomic_3m-1}) where large variations of this parameter have minor effects in the gapwidth, but in the other two families  (expression~\eqref{eq:perturbed_gap_heteroatomic_generic}) it is necessary to ensure the right energy ordering. 

\begin{figure}
    \centering
    \includegraphics[width=1.0\linewidth]{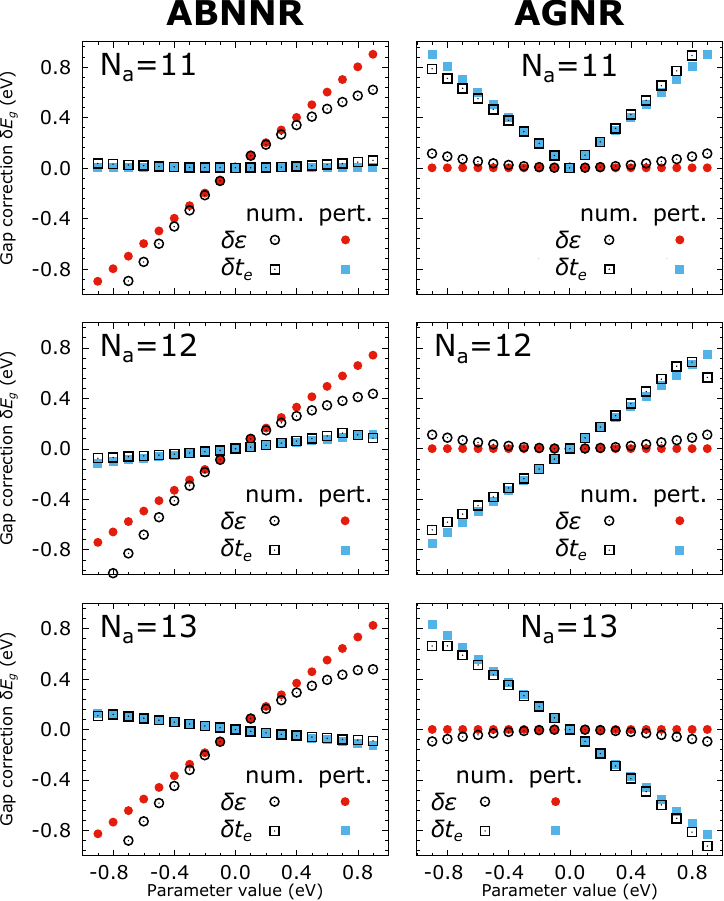}
    \caption{band gap correction $\delta E_g$ as a function of $\delta t_e$ (squares) and $\delta \epsilon$ (circles) in ABNNRs (left panels) and AGNRs (right panels) of width $N_a=11$, 12 and 13 from top to bottom.
    Colored symbols: perturbative formulae; Empty symbols: numerical diagonalization.}
    \label{fig:validity_perturbation}
\end{figure}

In addition we check the range of validity of the perturbative approach, by comparing the numerical solution (empty symbols in Figure~\ref{fig:validity_perturbation}) with the perturbative formulae (coloured symbols) in the range -1~eV, +1~eV in nanoribbons of width $N_a$=11, 12 and 13, i.e. one representative nanoribbon per family.
Globally, the numerical and the perturbative results are in very good agreement in the parameter range considered.
Deviations of $\delta \epsilon$ are a bit larger than those of $\delta t_e$ because numerical solutions display a quadratic trend whereas the perturbative formulae are linear in $\delta_\epsilon$.

\section{Conclusion}
We have carried out a comparative study of three gap engineering strategies on armchair BN and Gr nanoribbons, namely the application of an electric field, of strain and edge functionalization. 
The latter effect has been divided into two main contributions scrutinized separately: the electrostatic contribution and the deformation of the edges.
The study has been conducted by means of DFT calculations and by extending to the heteroatomic case a ladder model which we solved both numerically and perturbatively.

We have shown that the two materials, despite their similarities, respond in opposite and complementary ways to the different modifications they undergo.
In particular, while the gap of ABNNRs is much more sensitive to the application of electric field and to the electrostatic component of the edge functionalization, the gap of AGNRs is much strongly modified by the application of stress, being it global or localized on the edges. 

The heteroatomic model provides the explanation of this opposite behaviour.
In fact, the states responsible of the gap formation have very different symmetries in the two materials, being basically pure N or B states in ABNNRs and symmetric or antisymmetric combinations in AGNRs.
As a consequence, in ABNNRs the wave functions of the top valence and the bottom conduction are localised on the atomic sites which makes them much more sensitive to variations of the on-site energies (i.e. electrostatics).
On the contrary, in AGNRs the wave functions extend over the bonding between C atoms causing a much stronger sensitivity to variations of the hopping parameter (i.e. stress).

\begin{acknowledgments}
The authors acknowledge funding from the European Union’s Horizon 2020 research and innovation program under grand agreement N◦ 881603 (Graphene Flagship core 3) and from public grants overseen by the French National Research Agency (ANR) under the EXCIPLINT project (Grant No. ANR-21-CE09-0016).
\end{acknowledgments}

\section*{Appendix A: Computational details}
All DFT calculations are carried out within the generalized gradient approximation using the Perdew-Burke-Ernzerhof~\cite{Perdew1996} exchange correlation potential (PBE) as implemented in the Quantum ESPRESSO~\cite{Giannozzi_2009} simulation package.
To avoid interactions between consecutive cells, we include 15~\AA{ }and 20~\AA{ }of empty space in the $z$ and $x$ directions respectively.
In electron density calculations and relaxation runs, the periodic axis is sampled with 20 k-points centered in $\Gamma$. 
This mesh is dense enough to converge total energies in the smallest nanoribbons.
We use norm-conserving pseudopotentials~\cite{Hamann_2013} and set the kinetic energy cutoff at 80 Ry in both materials. 
It is worth stressing that using a large vertical empty space and a high energy cutoff is essential even in the relaxation runs in order to prevent nearly free-electron states from hanging below the $p_z$ states hence jeopardizing the gap description.
In fact, as already well known for free-standing layers~\cite{Posternak1983, Posternak1984,Blase1995, Paleari2019,Latil2023} and nanotubes~\cite{Blase1994a,Blase1994b,Hu2010} in BN nanomaterials there is a competition at the bottom conduction between $2p_z$ and $3s$ states, whose right alignment requires a dedicated convergence study. 
If sometimes one can overlook this issue in BN layers, because the two competing states create two separate valleys in the Brillouin zone, this is not the case in ABNNRs where both states give rise to a direct gap at $\Gamma$.

All reference structures are fully relaxed allowing optimization of both atomic positions and cell parameter $a$.
Relaxation runs have been performed with the Broyden–Fletcher–Goldfarb–Shanno (BFGS) algorithm with the stopping criterion of all forces being lower than $5 \times 10^{-5}$~eV/\AA.
In modified systems (i.e. calculations with applied electric fields, applied global or edge stress and with H displacements) no additional relaxation has been done.


\section*{Appendix B: Notable trigonometric identities}
\label{sec:appendix_trigonometric}
Some notable trigonometric identities used in the derivation of the TB perturbative corrections:
\begin{align}
\sin^2( \theta_n) + \sin^2(N_a \theta_n) &= 2 \sin^2( \theta_n)\,; \label{eq:equality_trigonometric_1}    \\
\sum_{j=1}^{N_a} \sin^2(j\theta_n) &= \frac{N_a+1}{2}\,; \label{eq:equality_trigonometric_2} \\
\sum_{j=1}^{N_a-1} \sin(j \theta_n)\sin\left[(j+1) \theta_n\right]  &= \frac{N_a + 1}{2} \cos(\theta_n)\,; \label{eq:equality_trigonometric_3}\\
\sum_{j=1}^{N_a} j \sin^2 \left( j \theta_n \right) &=\left( \frac{N_a+1}{2}\right)^2\,. \label{eq:equality_trigonometric_4}    
\end{align}


%

\end{document}